\documentclass[prl,showpacs,twocolumn]{revtex4}

\usepackage{graphicx}
\usepackage{amsmath}
\usepackage{slashed}
\usepackage{feynmp}

\begin{document}
\title{Soft mass generation}
\author{Ji\v r\'{\i} Ho\v sek}
\email{hosek@ujf.cas.cz} \affiliation{Department of Theoretical
Physics, Nuclear Physics Institute, Academy of Sciences of the Czech
Republic, 25068 \v Re\v z (Prague), Czech Republic}

\begin{abstract}
We replace the Higgs sector of the electroweak gauge $SU(2)_L \times
U(1)_Y$ model of three fermion families with its 'twenty-some'
parameters by a horizontal non-vector-like gauge $SU(3)_F$ quantum
flavor dynamics with one parameter. With plausible physical
assumptions we suggest that the new dynamics generates spontaneously
the masses of its eight flavor gluons, of leptons and quarks, and of
the intermediate $W$ and $Z$ bosons. Absence of axial anomalies
requires neutrino right-handed electroweak singlets and the dynamics
then suggests the existence of massive Majorana neutrinos.
\end{abstract}

\pacs{11.15.Ex, 12.15.Ff, 12.60.Fr}

\maketitle \noindent I. The Higgs mechanism \cite{Higgs} of soft
(i.e. spontaneous) mass generation in the Standard model \cite{GWS}
is built up on two basic principles: the principle of gauge
invariance and the principle of spontaneous breakdown of symmetry.
It has its roots in nonrelativistic field theory of quantum fluids
where both principles are instrumental for physical understanding of
many distinct macroscopic quantum phenomena. Soft mass generation is
in fact a necessity: Hard fermion and intermediate boson masses
simply ruin the unitary behaviour of scattering amplitudes with
longitudinally polarized intermediate vector bosons.

Principles are, however, more general than their particular
realizations: "...who has ever heard of a fundamental theory that
requires twenty-some parameters?" \cite{Lee} We think the Higgs
mechanism is a phenomenological realization of the principle of
spontaneous gauge symmetry breakdown in electroweak interactions in
much the same way the Ginzburg-Landau theory is a phenomenological
realization of the same principle in superconductivity.

Phenomenological interpretation of the Higgs mechanism means that
the massive spinless particle of a scalar field with properties
given by the Standard model Lagrangian does not exist.

What is then the 'microscopic' dynamics which generates softly the
vastly different masses to the quanta of three electroweakly
identical families of massless lepton and quark fields and to the
massless $W$ and $Z$ gauge fields? Attempts are numerous
\cite{DEWSB}. When strongly coupled they are not truly quantitative.
Our suggestion belongs to this category. For dynamical mass
generation we suggest to gauge properly the flavor index. Resulting
is {\it the strong horizontal non-vector-like non-confining
$SU(3)_F$ gauge quantum flavor dynamics} (QFD). It is defined by its
eight "phonons" or eight flavor gluons $C^{\mu}_a$ interacting
uniquely with each other and with leptons and quarks of both
chiralities with one coupling constant $h$.

In perturbation theory the masslessness of fermion fields is
protected by chiral symmetry and the masslessness of the gauge
fields is protected by gauge symmetry. {\it Massless fields can,
however, describe massive particles.} This is possible if: (1) The
nonlinear Schwinger-Dyson (SD) equations for the chirality-changing
fermion proper self-energies $\Sigma$ have energetically favorable
non-perturbative symmetry-breaking ultraviolet-finite solutions. (2)
In the transverse gauge-field polarization tensor
\begin{equation}
\Pi^{\mu\nu}_{ab}(q)\equiv
(q^2g^{\mu\nu}-q^{\mu}q^{\nu})\Pi_{ab}(q^2) \label{Pimn}
\end{equation}
the scalars $\Pi_{ab}$ develop dynamically the massless poles.

We argue as follows. First, $\Pi_{ab}$ develop dynamically the
massless poles. They correspond to  the eight composite 'would-be'
Nambu-Goldstone (NG) bosons of spontaneously broken global $SU(3)$
underlying QFD due to the flavor gluon self-interactions \cite{EF}
and the flavor gluon interactions with fermions \cite{JS}. This
realization of the general Schwinger mechanism \cite{Schwinger} is
theoretically viable if the underlying interaction is strong at
large distances, asymptotically free at small ones \cite{GW}, and
not vector-like \cite{VW}. Field-theoretic purity then demands the
model to be free of axial anomalies \cite{GJ}. Phenomenologically,
mediating the flavor changing, electric charge conserving processes
the flavor gluons have to be rather heavy. For definitess we
consider $M_a \sim 10^6 {\rm GeV}$.

Second, interactions of massive flavor gluons with both left-handed
and right-handed massless lepton and quark fields can build up the
bridges between these two in the form of the fermion symmetry
breaking proper matrix self-energies $\Sigma(q^2)$. Basically, the
charged lepton and quark masses of three electroweakly identical
fermion families differ due to a unique assignment of the the chiral
fermion multiplets to triplet and antitriplet representations of
QFD. Within the given electric charge the fermion masses differ due
to the low-momentum effective sliding coupling depending upon the
flavor gluon mass matrix. The prototype mass formula \cite{Pagels1}
$m_f=M \exp [-8\pi^2/h_{f}^2]$ nicely illustrates our task: If we
set $M=10^6\ {\rm GeV}$  the "neutrino" mass $m_{\nu} =10^{-9}\ {\rm
GeV}$ is obtained with an effective interaction strength
$h_{\nu}^2/4\pi=2\pi/15\ln10$, and the "top quark" mass $m_t=10^2\
{\rm GeV}$ with $h_{t}^2/4\pi=2\pi/4\ln10$. We will demonstrate that
due to the expected {\it non-analytic dependence of the
symmetry-breaking order parameters upon the effective interaction
strengths} such an output is conceivable.

Third, there is no mass-generating dynamics in electroweak $SU(2)_L
\times U(1)_Y$ interactions (small coupling constants $g$ and $g'$).
The fermion proper self-energies imply, however, also the
spontaneous breakdown of the 'vertical' electroweak symmetry.
Consequently, the Schwinger mechanism applies and the standard
analysis  of the corresponding Ward identities
\cite{HM},\cite{Hosek1},\cite{Hosek2} results in masses of $W$ and
$Z$ bosons expressed in terms of the fermion proper self-energies by
sum rules.
\\

\noindent II. Standard model chiral fermions $q_{fL}^{T}=(u_{fL},
d_{fL})$; $u_{fR}$; $d_{fR}$; $l_{fL}^{T}=(\nu_{fL}, e_{fL})$;
$e_{fR}$ of three families ($f=1,2,3$) can transform under $SU(3)_F$
either as triplets or anti-triplets. Non-vector-like assignments are
those in which not all fermion currents coupled to $C^{\mu}_a$ are
vectorial.

(i) Assume that $q_L$ is an $SU(3)_F$ triplet (i.e. both $u_L$ and
$d_L$ are triplets). Then $(u_R,d_R)$ can be either $(3,\bar 3)$ or
$(\bar 3,3)$, since for the choices $(3,3)$ and $(\bar 3,\bar 3)$
the mass matrices of the $u$- and $d$-type quarks would come out
equal. Without lack of generality choose $(u_R,d_R)=(3,\bar 3)$.

(ii) It follows that $l_L$ (i.e. both $\nu_L$ and $e_L$) cannot be a
triplet. For if it were, the charged lepton mass matrix would be
equal either to the $u$-type or the $d$-type quark matrix. Hence,
$l_L$ (i.e. both $\nu_L$ and $e_L$) must be an antitriplet.

(iii) Let $e_R$ be a triplet (case I). At this point we impose the
second restriction i.e., the absence of axial anomalies. Anomaly
freedom in this case requires introduction of {\it three neutrino
right-handed flavor triplets}, $\nu_{NR}$, $N=1, 2, 3$.

(iv) Let $e_R$ be an antitriplet (case II). Anomaly freedom in this
case requires introduction of {\it five neutrino right-handed flavor
triplets}, $\nu_{NR}$, $N=1,...,5$.

Two asymptotically free cases (three or five QFD triplets) of the
neutrino right-handed electroweak singlets should not be considered
as an ambiguity. Knowledge of the solution of the SD equations for,
say, the charged fermion masses would fix the neutrino pattern
uniquely.
\\

The QFD generates the full flavor gluon polarization tensor
$\Pi^{\mu\nu}_{ab}(q)$ (\ref{Pimn}). It must be symmetric in
flavor-octet indices by definition and transverse due to the
non-Abelian Ward identity. If $\Pi_{ab}(q^2)$ is proportional to
$\delta_{ab}$, the $SU(3)_F$ remains unbroken. Terms of the form
$\delta_{ab}\Pi^{(1)}_b(q^2)+d_{abc}\Pi^{(2)}_c(q^2)$ signal the
spontaneous breakdown of this symmetry. This is what we assume.
$\Pi^{\mu\nu}_{ab}(q)$ defines the full flavor gluon propagator
$\Delta^{\mu\nu}_{ab}(q)$ (for definiteness written in the
transverse Landau gauge):
\begin{equation}
\Delta^{\mu\nu}_{ab}(q)\equiv
\frac{-g^{\mu\nu}+q^{\mu}q^{\nu}/q^2}{q^2}[(1+\Pi(q^2))^{-1}]_{ab}
\end{equation}

We also assume that the QFD generates the fermion symmetry breaking
proper self energies $\Sigma$ which give rise to the fermion masses.
Later we find the symmetry breaking $\Pi$ and $\Sigma$
self-consistently. The full inverse fermion propagator $S(p)^{-1}$
is in general a three by three matrix in the flavor space
\cite{Benes}: $S(p)^{-1}=\slashed p-\hat \Sigma(p^2)$ where $\hat
\Sigma=\Sigma P_L+\Sigma^{+}P_R$ and
$P_{L,R}=\tfrac{1}{2}(1\mp\gamma_5)$. Such a propagator can be
explicitly inverted:
\begin{equation*}
S(p)=(\slashed p+\Sigma^{+})(p^2-\Sigma\Sigma^{+})^{-1}P_L+(\slashed
p+\Sigma)(p^2-\Sigma^{+}\Sigma)^{-1}P_R
\end{equation*}
{\it Massiveness of particles is a strong coupling low momentum
phenomenon:} The fermion and flavor gluon symmetry breaking
self-energies become important at low $q^2$. At high $q^2$ both
$\Sigma(p^2)$ and $\Pi(q^2)$ simply acquire their known symmetric
perturbative form.
\\

\noindent III. {\it Flavor gluon mass generation.} Here we follow
the analysis of the Ward identities with flavor gluons in accordance
with \cite{EF}: Divergences of the full vertices
$\Gamma^{\mu\nu\lambda}_{abc}(p+q,p)$ (three-flavor-gluon vertex)
and $\Gamma^{f;\mu}_{ij;c}(p+q,p)$ (fermion-flavor-gluon vertex) at
vanishing momenta are expressed in terms of the full inverse flavor
gluon and fermion propagators i.e., in terms of $\Pi$s and
$\Sigma$s, respectively. We assume that the ghost propagators do not
play any dynamical role in the generically nonperturbative
reasoning. This assumption is manifest in the 'pinch technique'
\cite{PT}. If the symmetry is unbroken the Ward identities are
fulfilled trivially. {\it If $\Pi_{ab}$ and $\Sigma$s develop the
symmetry breaking parts the validity of the Ward identities requires
the massless poles in the vertices themselves.} They correspond to
the 'would-be' NG bosons composed by construction from both flavor
gluons and from all fermion species in the world:
\begin{equation}
\Gamma^{\mu\nu\lambda}_{abc}(p+q,p)|_{pole}=
P_{bc;d}^{\nu\lambda}(p+q,p)\frac{i}{q^2}h(-iq^{\mu})\Lambda_{da}(q^2)\label{P}
\end{equation}\\
\begin{equation}
\Gamma^{f;\mu}_{ij;a}(p+q,p)|_{pole}=P^{f}_{ij;d}(p+q,p)\frac{i}{q^2}h
(-iq^{\mu})\Lambda_{da}(q^2)\label{P^f}
\end{equation}
where
\begin{equation}
-iq^{\mu}\Lambda_{da}(q^2)\equiv
[I^{\mu}_{C;da}(q)+\sum_{f}I^{\mu}_{f;da}(q)] \label{CNG}
\end{equation}
Physical interpretation of this decomposition should be clear:
\noindent (1) There are eight 'would-be' NG bosons composed both of
the flavor gluons and of all fermions in the model. \noindent (2)
$P^{\nu\lambda}_{bc;d}$ is the effective coupling of the NG boson
with flavor gluons. \noindent (3) $P^f_{ij;d}$ is the effective
coupling of the NG boson with the fermion $f$. \noindent (4)
$I^{\mu}_{C;da}(q)$ and $I^{\mu}_{f;da}(q)$ are the vectorial
tadpole UV finite loop integrals. They convert in terms of the
effective vertices $P$, the elementary vertices and the full flavor
gluon and fermion propagators both the flavor gluon components and
the fermion components of the 'would-be' NG bosons to the flavor
gluons. \noindent (5) The crucial effective bilinear derivative
vertex between the flavor gluon octet and the 'would-be' NG boson
octet is given by (\ref{CNG}).

The vertex (\ref{CNG}) gives rise to the massless pole in the
longitudinal part of the flavor gluon polarization tensor
(\ref{Pimn}). Although its transversality is saved by contributions
which we cannot compute explicitly, it follows from it that the
flavor gluon mass matrix $M^2_{ab}(q^2)$ is given by the formula
\cite{EF}
\begin{equation}
-q^2\Pi_{ab}(q^2)\equiv
M^2_{ab}(q^2)=\sum_{d}\Lambda_{ad}(q^2)\Lambda_{bd}(q^2) \label{Mg}
\end{equation}
Practical applications will demand diagonalization of the mass
matrix $M^2_{ab}(0)$ and introduction of the flavor gluon mass
eigenstates.
\\

\begin{figure}[h]
\begin{center}
\includegraphics[width=0.45\textwidth]{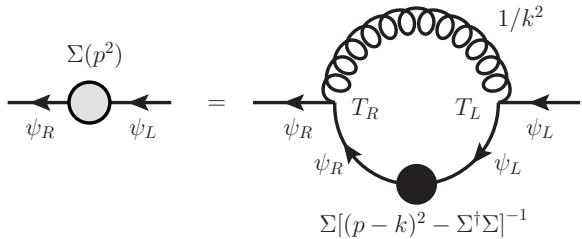}
\caption[]{Structure of the SD equation for chirality changing
$\Sigma$ of a fermion (charged lepton or quark) $\psi$. $T_{L,R}$
are the triplet ($\frac{1}{2}\lambda$) or antitriplet
($-\frac{1}{2}\lambda^*$) generators of a given chiral fermion
$\psi_{L,R}$.} \label{obrazek}
\end{center}
\end{figure}

\noindent IV. {\it Fermion mass generation.} Structure of the SD
equations for the chiral symmetry changing fermion proper self
energies $\Sigma(p^2)$ (NJL type self-consistency condition
\cite{NJL}) of electrically charged fermions is easily read off the
interaction Lagrangian of QFD using the general form of the massive
fermion propagator. It is shown in Fig.\ref{obrazek}. The (bare)
flavor gluon propagator is taken in the Feynman gauge as suggested
by the pinch technique \cite{PT}.

The neutrino SD equation is more subtle due to possible Majorana
mass terms. Here we merely point out that for three generations the
hard Majorana mass terms are prohibited by symmetry and the complete
neutrino self energy (Dirac plus Majorana) must be generated
dynamically. Such a work is in progress.

The integration in Fig.1. extends over all momenta and the SD
equations {\it must} be improved by taking into account properly the
momentum-dependent sliding coupling $\bar h^2_{ab}(k^2)$
\cite{Pagels1}. We know no way of knowing $\bar h^2_{ab}(k^2)$ at
low momenta other than solving the theory \cite{wilson}. It is
likely that it is dominated by the exchanges of the composite
'would-be' NG bosons with the effective vertices
$P_{bc;d}^{\nu\lambda}(p+q,p)$ and $P^{f}_{ij;d}(p+q,p)$ to both
gluons and fermions, respectively.

To proceed we write
\begin{equation}
\frac{1}{k^2}=\frac{1}{k^2}\{[1+\Pi(k^2)]^{-1}+\Pi(k^2)[1+\Pi(k^2)]^{-1}\}
\label{Pi}
\end{equation}
and argue as follows:

(1) At high momenta $\Pi$ is given by perturbation theory.  The
first term in (\ref{Pi}) when used in Fig.1. gives rise to flavor
insensitive $\Sigma$ due to massless flavor gluon exchange with
asymptotically free flavor insensitive interaction strength $\bar
h^2(k^2)= h^2[1+\Pi(k^2)]^{-1}$. The corresponding SD equation
giving the high-momentum asymptotics of $\Sigma$ was studied in QCD
in detail in \cite{Pagels2}. The second term in (\ref{Pi})
corresponding to massless gluon exchange with the bare charge should
be ignored.

(2) At low momenta the non-perturbative $\Pi$ is given by
(\ref{Mg}). The first term in (\ref{Pi}) corresponds to the massive
gluon exchanges with a bare charge
($\tfrac{1}{k^2}(1+\Pi)^{-1}=(k^2-M^2)^{-1}$) and in Fig.1. it
should be ignored. The second term in (\ref{Pi}) when used in Fig.1.
gives rise to $\Sigma$s due to massless gluon exchange with the
low-momentum $\bar h^2_{ab}(k^2)$ running to a non-perturbative IR
stable fixed point:
\begin{equation}
\bar h^2_{ab}(k^2)=h_{*}^2[\Pi(k^2)(1+\Pi(k^2))^{-1}]_{ab}
\label{hIR}
\end{equation}
The corresponding matrix SD equations, still rather schematic,
hopefully illustrate the main point: The symmetry-breaking form of
$\Pi_{ab}$ implies that {\it at low momenta} the fermion  self
energies $\Sigma$ differ in different flavor channels by the
low-momentum flavor sensitive interaction strengths (\ref{hIR}),
basically due to the low momentum symmetry breaking flavor gluon
self energy.

At this exploratory stage we are merely able to illustrate that the
low momentum Ansatz (\ref{hIR}) for the sliding coupling is bona
fide responsible for strong suppression of the fermion mass with
respect to the huge flavor gluon mass. We replace both
$M^2_{ab}(k^2)$ and the fermion self energies $\Sigma_{ij}(p^2)$
=$\Sigma_{ij}(0)=m_{ij}$ by real numbers $M^2$ and $m$,
respectively. The SD equation in Fig.1. turns into an algebraic
equation
\begin{equation}
m=\frac{h_{*}^2}{16\pi^2}\int_{0}^{\infty}dk^2\frac{M^2}{k^2+M^2}
\frac{m}{k^2+m^{2}} \label{m}
\end{equation}
with solution $ m=M\exp[-8\pi^2/h^{*2}]$ announced earlier in the
paper.

Finding reliable low momentum dominated matrix symmetry breaking
self-consistent fermion and flavor gluon self-energies which define
the fermion and the flavor gluon mass spectrum is an exceedingly
difficult task for future work. Life with nonperturbative QCD taught
us, however, to be meek.
\\

\noindent IV. {\it Masses of $W$ and $Z$ bosons} are the necessary
consequence of the dynamically generated fermion masses: The fermion
proper self-energies $\Sigma(p^2)$ generated by strong QFD break
spontaneously also the 'vertical' $SU(2)_L\times U(1)_Y$ symmetry
down to $U(1)_{em}$. Consequently, as before, the properties of
three composite 'would-be' NG bosons can be extracted from the
$SU(2)_L\times U(1)_Y$ Ward identities
\cite{HM},\cite{Hosek1},\cite{Hosek2}. For simplicity we consider
the fermion proper self energies diagonal.
\begin{multline*}
\Gamma^{\alpha}_{W}(p+q,p)=\frac{g}{2\sqrt2}\{\gamma^{\alpha}(1-\gamma_5)-\\
-\frac{q^{\alpha}}{q^2}[(1-\gamma_5)\Sigma_U(p+q)-(1+\gamma_5)\Sigma_D(p)]\},
\end{multline*}
\begin{multline*}
\Gamma^{\alpha}_{Z}(p+q,p)=\frac{g}{2\cos\theta_W}\{t_3\gamma^{\alpha}(1-\gamma_5)-\\
-2Q\gamma^{\alpha}\sin^2\theta_W-\frac{q^{\alpha}}{q^2}t_3
[\Sigma(p+q)+\Sigma(p)]\gamma_5\}.
\end{multline*}
When the electroweak gauge interactions are switched on as weak
external perturbations, the $W$ and $Z$ bosons dynamically acquire
masses. Their squares are defined as the residues at single massless
poles of the $W$ and $Z$ polarization tensors:
\begin{equation}
m_W^2=\frac{1}{4}g^2\sum(m_U^2 I_{U;D}(0)+ m_D^2 I_{D;U}(0))
\label{mW}
\end{equation}
\begin{equation}
m_Z^2=\frac{1}{4}(g^2+g'^2)\sum(m_U^2 I_{U;U}(0)+ m_D^2 I_{D;D}(0))
\label{mZ}
\end{equation}
In the formulas above $U$ and $D$ abbreviate upper and nether
fermions in electroweak doublets, respectively. The neutrinos are
considered as massive Dirac fermions for simplicity. The quantities
$I$ in (\ref{mW}, \ref{mZ}) defined in \cite{Hosek1} are the UV
finite loop integrals depending upon $\Sigma$s. If the proper
self-energies $\Sigma_U$ and $\Sigma_D$ were degenerate the Weinberg
relation $m_W^2/m_Z^2\cos^2\theta_W=1$ would be fulfilled.
Quantitative analysis of departure from this relation demands
quantitative knowledge of the functional form of proper
self-energies. At present we can only refer to an illustrative model
analysis of \cite{Hosek1}.
\\

\noindent V. Present model of soft generation of masses of the
Standard model particles by a strong-coupling dynamics, having just
one unknown parameter $h$, is either right or plainly wrong.
Reliable computation of the fermion mass spectrum is, however, far
away. Ultimately, masses should be related. (1) One elaborated
example of mass relations is the sum rules for the intermediate
boson masses $m_W$ (\ref{mW}) and $m_Z$ (\ref{mZ}). The implication
is interesting: {\it There is no generic Fermi scale in the model}.
The intermediate boson masses are merely a manifestation of the
large top quark mass \cite{Hosek1},\cite{Hosek2}.  (2) Detailed
analysis of the uniquely defined neutrino sector is a challenge. The
very existence of sterile neutrinos introduced for anomaly freedom
should have experimental consequences in neutrino oscillations and
in astrophysics \cite{kus}. (3) The fermion SD equations can fix
also the fermion mixing parameters. (4) It is natural to expect that
the unitarization of the scattering amplitudes with longitudinal
polarization states of massive spin one particles proceeds in the
present model via the massive composite 'cousins' of the composite
'would-be' NG bosons. Its practical implementation is obscured,
however, by our ignorance of the detailed properties of the spectrum
of strongly coupled $SU(3)_F$.

In conclusion we may perhaps defend ourselves by paraphrasing the
godfather of the Higgs mechanism, Fritz London \cite{London}: 'the
model at which we have arrived is distinguished by its uniqueness in
such a way that we could hardly avoid writing it down'.

\begin{acknowledgments}
I am grateful to Petr Bene\v s, Tom\'a\v s Brauner and Adam Smetana
for help with the manuscript and for discussions. Financial support
of this work by the Committee for the CR-CERN collaboration is
gratefully acknowledged.
\end{acknowledgments}

\end{document}